# HIGH STRAIN RATE COMPRESSION OF EPOXY MICROPILLARS

Mario Rueda-Ruiz[1,2], Miguel A. Monclus[1], Ben D. Beake[3], Francisco Gálvez[2] and Jon M. Molina-Aldareguia[1*]

[1]IMDEA Materials Institute, C/Eric Kandel 2, 28906 Getafe, Madrid, Spain

[2]Department of Materials Science, Universidad Politécnica de Madrid, E. T. S. de Ingenieros de Caminos, 28040 Madrid, Spain

[3]Micro Materials Ltd, Willow House, Yale Business Village, Ellice Way, Wrexham LL13 7YL, UK

**Abstract**

The study of the high strain rate mechanical behaviour of materials using micropillar compression tests has been hindered so far due to the lack of suitable instrumentation. In the present study, a novel high strain rate micropillar compression set-up is introduced. The high speed tests are carried out with the impact configuration of a pendulum-based nanoindentation device that was instrumented with force-sensing capability by means of a piezoelectric load cell. This is necessary due to the considerable inertia effects that make the actual applied force on the sample much higher than the actuated one. The proposed novel technique was successfully applied to study the mechanical behaviour of an epoxy resin over a wide range of strain rates. The results reveal a significant strain rate sensitivity that correlates well with that obtained for the same material at the macroscale, validating the novel set-up.

## 1. INTRODUCTION

The mechanical characterization of materials at the micro- and nano-scales is becoming an important discipline in engineering and science. This is driven, partly, by recent advances in fields like microelectromechanical systems (MEMS) and advanced coatings, and the need to understand the physics behind the size-dependent deformation of materials. And also, by the need to develop micromechanical tests that allow determining the mechanical properties of the constituents in bulk composite materials, specially towards incorporating them into multiscale models of composite behavior [1]. Nanoindentation has been traditionally the main tool for nano-/micromechanical characterization of materials, not only for the determination of hardness and elastic modulus [2], but also for extracting the plastic constitutive behaviour [3,4]. However, the determination of the constitutive behaviour has proven to be especially challenging due to the large strain gradients and complex stress states that develop in the material during indentation. This issue was overcome with the pioneering work of Uchic et al. who demonstrated the possibility of fabricating micron-size pillars in a material surface that could be tested with a conventional nanoindentation system [5,6]. This way, the stress and strain are uniform along the tested material and the uniaxial constitutive behaviour can be directly probed, without the need of a complex inverse analysis.

---

[*] Corresponding author: jon.molina@imdea.org; IMDEA Materials Institute, C/ Eric Kandel 2, 28906 Getafe, Madrid, Spain

Since then, the micropillar compression technique has been applied to study the mechanical behaviour of all classes of materials: from the study of size-dependent deformation in metals [7,8], ceramics [9,10] and polymers [11–13], to the calibration of micromechanical models of polycrystalline metals [14] and fibre-reinforced polymer composites [15–17].

Micropillar compression tests have been also performed over a wide range on environments and loading conditions, following new developments in nanoindentation instrumentation, such as high temperature and varying strain rates [18,19]. A large number of studies have dealt with the study of high temperature deformation [20,21], and more recently with material behaviour under cryogenic conditions [22]. In terms of loading conditions, micropillar compression tests have been performed under conditions of high cycle fatigue [23] and creep [24]. An area that has received less attention is the study of high strain rate behaviour, due to the absence of suitable instrumentation for obtaining reliable data. Understanding how materials deform under conditions of high strain rates is of great importance to model machining and forming processes, and materials subjected to impact events. Furthermore, the high strain rate characterization at the microscale allows extending the *in situ* calibration of micromechanical models of materials, like crystal plasticity models, to higher strain rates. Recently, Guillonneau et al. [25] demonstrated the capabilities of a novel nanoindentation device to perform high strain rate micropillar compression, and applied the technique to study the rate dependent deformation behaviour of nanocrystalline nickel. The technique was later expanded to the study of ceramics materials [26] and polymers [27]. The high strain rate characterization of polymers is of particular importance because these materials can be highly rate sensitive [28]. New developments in this area will open the door to the calibration of micromechanical models of the impact behaviour of fibre-reinforced polymer composites, incorporating strain rate dependent behaviour, a topic that has remained elusive so far [15].

In the present study, a new set-up for performing high strain rate micropillar compression tests that offers reliable data is presented. The set-up is based on impact indentation and it uses a pendulum-based nanoindentation device modified to measure force via a piezoelectric sensor. Micropillar compression tests over a wide range of strain rates were performed on an epoxy resin typically used as matrix material in fibre-reinforced polymer composites. Results were validated against macroscopic mechanical characterization results obtained in the same material, using a range of devices, including a electromechanical testing machine, a hydraulic testing device and a split-Hopkinson pressure bar (SHPB). Unlike the macroscopic tests, that require access to bulk resin specimens, the impact micropillar compression test allows the *in situ* testing of the resin matrix within the real composite material, so this test unravels the capability of calibrating composite micromechanical models at high strain rates.

## 2. EXPERIMENTAL PROCEDURES

### 2.1 Material and specimen preparation

The tested material was an aerospace-grade Hexcel 8552® epoxy resin, used as matrix material in fibre-reinforced composite materials. The resin in fresh state was heated up to 100 °C for 30 mins and speed-mixed at 2300 rpm for 1 min to eliminate entrapped air. It was then poured into open moulds and cured in an autoclave for 135 mins at 180 °C with a heating/cooling rate of 2 °C/min and a pressure of 7 bar. The autoclave process ensured that porosity was kept to a minimum.

Specimens for micropillar compression were prepared by first cutting the plates of cured epoxy in smaller pieces of about 1x1x0.5 mm$^3$. The specimens were coated with a few nm thick gold layer to avoid charging effects during the fabrication of the micropillars. The micropillars were manufactured using a Helios NanoLab DualBeam 600i of Thermo Fisher Scientific (USA), equipped with a Focused Ion Beam (FIB). Cylindrical pillars were milled using Ga$^+$ ion beam and an annular milling strategy. The procedure was carried out in three steps of equal accelerating voltage (30 kV) and decreasing current. The first two steps used 21 nA and 9.3 nA to create a rough shape of the pillar. The last step was gas assisted milling, using $(MnSO_4) \cdot 7H_2O$ and a beam current of 0.43 nA to create the final shape of the pillar and to ensure a smooth surface. In total, 12 micropillars were carved in the epoxy surface, with a top diameter of around 7 µm, an aspect ratio of about 3 and a slight taper angle of 1-2°.

Specimens for macroscale compression were extracted from the same batch of plates of cured epoxy using computer numerical control (CNC) machining. Samples were produced with a square cross section of 5x5 mm$^2$ and a height of 10 mm. Sample geometry was selected following requirements for stress equilibrium of the high strain rate tests performed on the split-Hopkinson pressure bar.

## 2.2 Microscale mechanical characterization

Low and high speed micropillar compression tests were carried out on an instrumented nanoindenter NanoTest Alpha from Micro Materials Ltd. (Wrexham, UK), using a diamond flat punch with a diameter of 10 µm. This is a force-actuated, displacement-sensing pendulum-based nanoindentation device with the capability of performing low speed conventional nanoindentations, as well as high speed ones using a special configuration for performing nanoimpacts. In the low speed configuration, force is actuated via an electromagnetic coil and displacement is sensed using two capacitor plates in line with the indenter. A schematic of the set-up is shown in Figure 1. In the conventional configuration, the maximum reliable displacement rate is of the order of 100 nm/s, which for a pillar with a height of the order to 10 µm corresponds to a maximum strain rate of the order of $10^{-2}$ s$^{-1}$. In this work, the low strain rate micropillar compression tests were carried out at two low speed conditions, 10 and 100 nm/s, using a closed-loop displacement-controlled scheme up to a maximum displacement of 5 µm. 3 pillars were tested per condition.

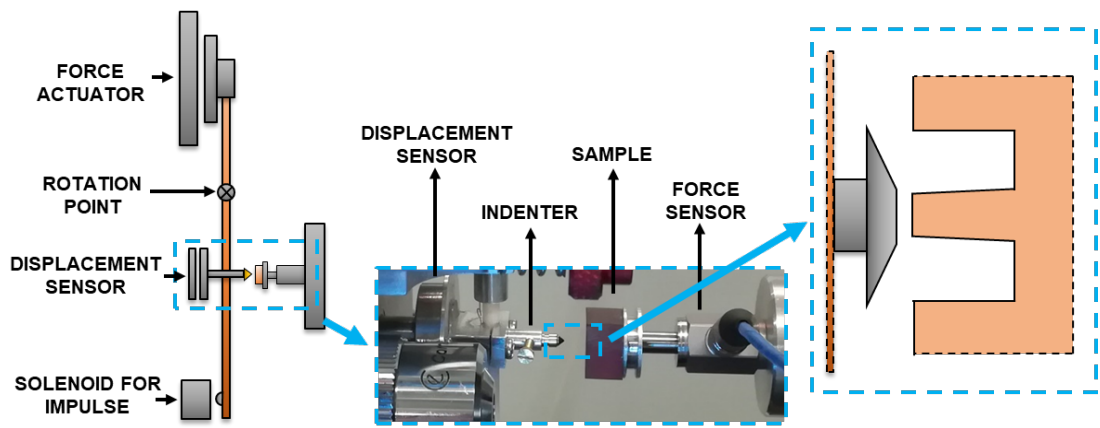

**Figure 1 – Schematics of the pendulum impactor and the force sensor added to the setup**

In the impact configuration, a material of low magnetic hysteresis attached to the bottom of the pendulum is attracted by a solenoid. Then, when the solenoid voltage is switched off, the indenter swings towards the material surface driven by the impulse force applied in the actuator. The impact set-up allows reaching displacement rates in the order of 1 mm/s, extending the potential maximum strain rate for a 10 µm high micropillar up to $10^2$ s$^{-1}$. However, the output of this impact setup is the displacement signal over time as the indenter penetrates the sample at large velocities. To measure the load, the device was instrumented for this work with an independent force sensing capability by using a force sensor in direct contact with the sample. This was necessary because of the significant contribution of inertia forces, that makes the applied force in the material much higher than the actuated force. A miniature piezoelectric force sensor ICP-209C11 of PCB

Piezotronics (USA) was selected because of its high frequency response (high stiffness) and its ability to measure in the mN range. The sample was glued using cyanoacrylate adhesive to an Al stub that was directly screwed to the force sensor. The setup was already used successfully to carry out nanoindentation tests at impact velocities [29]. For the high-strain rate micropillar compression tests, the impulse force was set to 2 and 5 mN and the impulse distance to 20 μm, which produced an initial impact velocity of the flat punch against the micropillar of 0.5 and 0.9 mm/s, respectively. In this case, the nanoimpact configuration does not allow to prescribe the maximum displacement, which was a result of the initial velocity and the micropillar response. 9 pillars were tested under impact conditions.

### 2.3 Macroscale mechanical characterization

Low speed testing was performed on an electromechanical test frame Instron 5966 (USA) instrumented with a 10 kN load cell. Strain was computed from bench displacement, accounting for frame compliance. Tests were carried out at three constant cross-head speeds of 4, 40 and 400 mm/min. Tests at medium strain rates were performed on a servo-hydraulic system Gleeble 3800 of Dynamic Systems (USA). Strain was computed from the machine stroke, accounting for frame compliance, and force was measured using a load cell. Tests were carried out at constant stroke rates of 100, 200 and 1000 mm/s. The high speed tests were performed on a split-Hopkinson pressure bar (SHPB) consisting of Inconel incident, transmitted and striker bars. The striker bar was launched using a gas gun. The incident and transmitted bars were instrumented with biaxial strain gauges. Stress and strains in the SHPB test were computed from the conventional 1D wave analysis using the strain signals from the incident and transmitted bars [28]. 5 specimens were tested per condition for statistical analysis.

### 2.4 Data reduction for micropillar compression tests

Stress and strain can be directly computed from the micropillar compression tests because they are uniform along the length of the pillar. True stress and strain were calculated assuming that the tested volume is kept constant:

$$\sigma_{true} = \frac{4P}{\pi D^2}\left(1 - \frac{u-u_b}{L}\right) \quad (1)$$

$$\epsilon_{true} = -ln\left(1 - \frac{u-u_b}{L}\right) \quad (2)$$

where $P$ is the applied force and $u$ is the total displacement measured at the top surface of the pillar. $u_b$ is the displacement of the pillar base and was estimated using Sneddon´s correction, which assumes elastic deformation of the base, $u_b = (1 - v_s^2)P/(E_s D)$ [30,31]. $E_s$ and $v_s$ are the sample's Young's modulus and Poisson's ratio, respectively, $L$ is the pillar length and $D$ is the pillar diameter. The top diameter is commonly used for the calculations; however, it has been observed that this leads to overestimations in the stress calculation in tapered pillars [31]. A set of finite element modelling (FEM) simulations of the micropillar compression test was performed to study the effect of taper in the estimated stress in order to correct for it, and the results can be found in the Supplementary Data file. It was observed that a taper angle ($\theta$) of 1-2°, as in the pillars manufactured in this study, led to significant deviations on the computed stress. This was corrected in this work by using an average diameter, computed as $D_t + Ltan\theta$, being $D_t$ the top diameter, an approach that has been followed before [12,32].

### 3. RESULTS AND DISCUSSION

Figure 2 plots the stress-strain curves and strain rates for the micropillar in all loading conditions and for all repeats. The strain rate achieved when using the impact configuration, of the order of 10-100 $s^{-1}$, was about four orders of magnitude larger than the strain rate obtained for the low speed cases. Besides, the strain rate was fairly constant throughout the strain range of interest. Even higher strain rates can potentially be achieved in the impact configuration by impacting the pillar with higher velocities; however, this was not feasible due to limitations in the data acquisition rate of the current configuration. In addition, the curves in Figure 2 show a significantly noisier response at high strain rates. This is because the measurement was performed near the sensitivity limit of the force sensor (the noise floor was of the order of ± 0.5 mN for the current setup), while in the low speed cases, the force is actuated and not measured. Nevertheless, the curves were of sufficient quality for the extraction of mechanical properties. Regarding the reproducibility of the results, there was a significant scatter in the stress-strain curves performed at the same strain rate. The scatter is inherent to any microscale testing technique and is consistent with the behaviour observed in other epoxy micropillars fabricated by FIB milling [11,17]. The major potential challenges to reduce the scatter of micropillar compression tests in a homogenous material like an epoxy resin are the control of misalignment during the compression tests and the precise measurement of the micropillar dimensions. Based on the FEM simulations, alignment errors are not the major source of the scatter show. As a matter of fact, it was found that scatter due to alignment errors is not very significant, of the order of 10%, for a realistic maximum misalignment angle between the flat punch axis and the pillar axis of 5°, as shown in the Supplementary Data file. On the contrary, the precise measurement of the micropillar dimensions remains experimentally elusive, due to the irregularities produced by FIB milling at the pillar base, it is difficult to measure the bottom diameter and the micropillar height precisely (see for instance the example of an as-prepared micropillar in Figure 5-A). Using lathe FIB milling instead of an annular strategy could potentially facilitate a more accurate description of the micropillar geometry; however this method is much more time consuming [17]. Other authors have proven an excellent reproducibility of results in polymer micropillars fabricated by other methods, like lithography [12] or by direct moulding of the polymer at the microscale [13]. However, the FIB process has the advantage of fabricating pillars *in situ* in selected locations within a desired microstructure, such as in the resin pockets of fibre-reinforced polymer composites [17], which cannot be done by any other microfabrication methods.

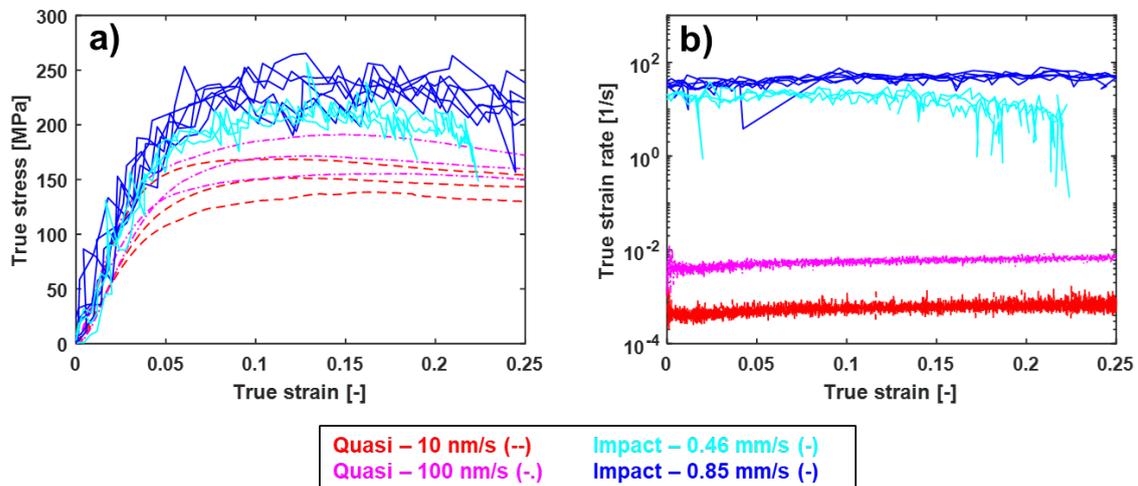

**Figure 2 – (a) Stress-strain curves from micropillar compression tests over a wide range of strain rates ($10^{-3}$ to $10^{2}$ $s^{-1}$) and (b) true strain rate registered throughout the test**

The stress-strain curves in Figure 2 exhibited the typical deformation behaviour of epoxy resins under compression loading. The first part was linear, corresponding to the elastic deformation of the material, up to the onset of plastic deformation. The deformation then became nonlinear and the stress increased up to the maximum compressive stress, followed by a slight strain softening. The epoxy specimens tested in compression at the macroscale exhibited the same behaviour, as shown in the stress-strain curves presented in Figure 3. Furthermore, the shape of the low speed curves was in good agreement with the work of Naya et al. where the same resin system Hexcel 8552® was tested under quasistatic conditions [17].

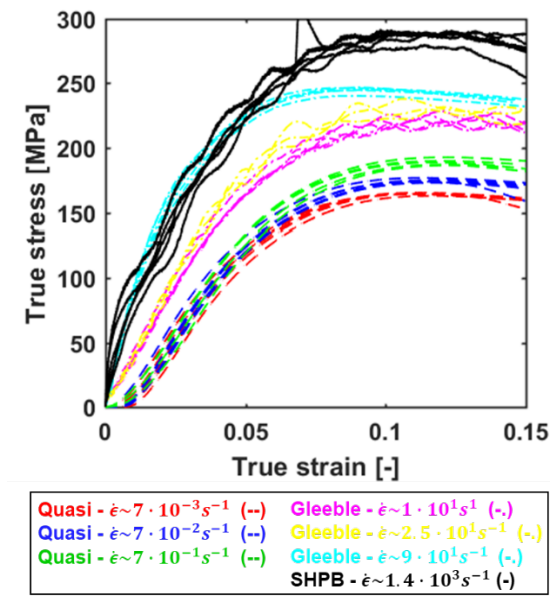

**Figure 3** – Stress-strain curves from macroscale compression tests performed at the following strain rates: 7·10⁻³, 7·10⁻², 7·10⁻¹, 1.5·10³ s⁻¹

The two important mechanical properties that can be extracted from the stress-strain curves obtained from the micropillar compression tests are the elastic modulus and the maximum compressive strength, which are presented in Figure 4 as a function of true strain rate. The elastic modulus was obtained by fitting a linear polynomial to the first part of the curve (in the strain range from 1% to 1.5%) whereas the strength was obtained as the average stress in the near-plateau region after the maximum stress (in the strain range from 10% to 20%). The mechanical properties obtained in the micropillar compression tests correlated well with those obtained from the macroscale compression tests. The maximum compressive stress obtained with the macroscale compression tests exhibits a variable strain rate sensitivity with strain rate. At quasi-static strain rates, the strain rate sensitivity was 0.03. However, the results at higher strain rates indicated a higher rate sensitivity of 0.05. The transition in strain rate sensitivity occurred at strain rates of around $10^{-1}$ s⁻¹. This bilinear behaviour has been reported before in epoxy resins and has been attributed to the effect of the molecular β-transition [33]. Beyond the β-transition, secondary molecular motions of the polymer chain are restricted and this induces a stronger response of the material at high strain rates. Regarding the elastic modulus, an increase would also be expected for the higher strain rates [28]. However, the micropillar results did not capture any substantial effect of strain rate in elastic modulus, while the macroscale data exhibits certain increase in the modulus.

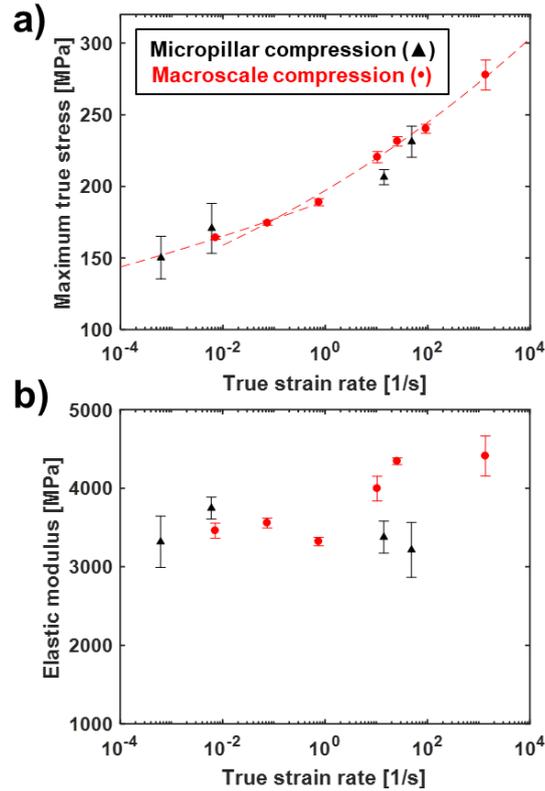

**Figure 4 – Comparison of maximum stress (a) and elastic modulus (b) versus true strain rate obtained from micropillar compression and macroscale compression**

The close correlation found between results obtained from both micropillar and macroscale compression tests (see Figure 4) raises the question about whether the comparison of mechanical properties at the two scales is valid due to size effects. Wang et al. studied the issue of size effects in epoxy resins and found that they only manifested at the submicron scale and attributed this to the lower probability of percolations of weak bonds in the epoxy molecular network [11]. Naya et al. studied size effects for micrometer size micropillars on the Hexcel 8552® epoxy system and found that the maximum compressive strength was significantly size dependent [17]. They attributed this to the confining effect of a stiff skin around the pillar formed due to Ga+ ion implantation during the FIB milling process. This stiff skin also appeared in the micropillars made in the present study, as revealed by the surface wrinkles that appeared on the micropillar surface after low speed deformation (red arrows in Figure 5-B). At high speeds, the skin spallation was not perceived because the maximum displacement could not be prescribed as explained before and therefore, the pillars were severely deformed, as shown in Figure 5-C. In any case, Naya et al. [17] determined that the skin-induced size effect only introduced size effects for pillar diameters below 6 μm. Since the diameter of the pillars manufactured in this study was around 7 μm, the current micropillar results are free of size effects, and hence, their close correlation with the macroscale SHPB tests obtained in Figure 4 constitutes a further experimental validation of the new instrumentation for impact micropillar compression presented in this work.

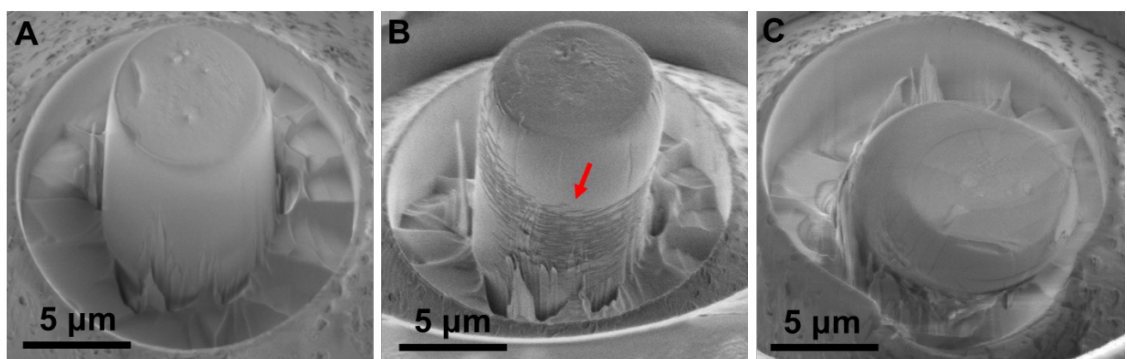

**Figure 5** - SEM images of a pillar before testing (a), the same pillar after low speed test (b) and a pillar after impact test (c)

## 4. CONCLUSIONS

A novel test set-up for high strain rate micropillar compression testing was proposed and applied to the study of the mechanical behaviour of an epoxy resin over a wide range of strain rates. The high strain rate test is based on impact loading, which is performed on a pendulum-based nanoindenter modified to measure the actual force imposed on the sample via a piezoelectric force sensor. The micropillar compression tests on the epoxy resin revealed a significant strain rate sensitivity of the maximum compressive strength that correlated well with the values extracted from macroscopic compression tests in a wide range of strain rates. A bilinear strain rate sensitivity was found, with a transition from 0.03 to 0.05 at a strain rate of around $10^{-1}$ $s^{-1}$, that is attributed to the contribution of two different mechanisms of molecular motion in the plastic deformation of epoxy resins with strain rate. All in all, the novel test set-up presented in this study provides a unique tool for the study of the high strain rate mechanical behaviour of all material types at the microscale.

**Acknowledgements**

The research leading to these results has received funding from the European Union's Horizon 2020 research and innovation programme under the Marie Sklodowska-Curie grant agreement Nº 722096, DYNACOMP project.

**Data availability statement**

The datasets generated during and/or analysed during the current study are available from the corresponding author on reasonable request.

**Competing interests statement**

The authors declare that they have no competing interests.